\documentstyle[12pt]{article}
\newcounter{numerone}
\newcounter{numertwo}
\newcounter{numerthree}
\newcounter{numerfive}
\expandafter\ifx\csname amssym.def\endcsname\relax \else\endinput\fi
\expandafter\edef\csname amssym.def\endcsname{%
       \catcode`\noexpand\@=\the\catcode`\@\space}
\catcode`\@=11
\def\undefine#1{\let#1\undefined}
\def\newsymbol#1#2#3#4#5{\let\next@\relax
 \ifnum#2=\@ne\let\next@\msafam@\else
 \ifnum#2=\tw@\let\next@\msbfam@\fi\fi
 \mathchardef#1="#3\next@#4#5}
\def\mathhexbox@#1#2#3{\relax
 \ifmmode\mathpalette{}{\m@th\mathchar"#1#2#3}%
 \else\leavevmode\hbox{$\m@th\mathchar"#1#2#3$}\fi}
\def\hexnumber@#1{\ifcase#1 0\or 1\or 2\or 3\or 4\or 5\or 6\or 7\or 8\or
 9\or A\or B\or C\or D\or E\or F\fi}

\edef\@scale{\ifcase\@ptsize \@m \or\magstephalf\or\magstep1\fi}

\font\tenmsa=msam10 scaled \@scale
\font\sevenmsa=msam7 scaled \@scale
\font\fivemsa=msam5 scaled \@scale
\newfam\msafam
\textfont\msafam=\tenmsa
\scriptfont\msafam=\sevenmsa
\scriptscriptfont\msafam=\fivemsa
\edef\msafam@{\hexnumber@\msafam}
\mathchardef\dabar@"0\msafam@39
\def\dashrightarrow{\mathrel{\dabar@\dabar@\mathchar"0\msafam@4B}}
\def\dashleftarrow{\mathrel{\mathchar"0\msafam@4C\dabar@\dabar@}}

\def\ulcorner{\delimiter"4\msafam@70\msafam@70 }
\def\urcorner{\delimiter"5\msafam@71\msafam@71 }
\def\llcorner{\delimiter"4\msafam@78\msafam@78 }
\def\lrcorner{\delimiter"5\msafam@79\msafam@79 }
\def\yen{{\mathhexbox@\msafam@55 }}
\def\checkmark{{\mathhexbox@\msafam@58 }}
\def\circledR{{\mathhexbox@\msafam@72 }}
\def\maltese{{\mathhexbox@\msafam@7A }}

\font\tenmsb=msbm10 scaled \@scale
\font\sevenmsb=msbm7 scaled \@scale
\font\fivemsb=msbm5  scaled \@scale
\newfam\msbfam
\textfont\msbfam=\tenmsb
\scriptfont\msbfam=\sevenmsb
\scriptscriptfont\msbfam=\fivemsb
\edef\msbfam@{\hexnumber@\msbfam}
\def\Bbb#1{{\fam\msbfam\relax#1}}
\def\widehat#1{\setbox\z@\hbox{$\m@th#1$}%
 \ifdim\wd\z@>\tw@ em\mathaccent"0\msbfam@5B{#1}%
 \else\mathaccent"0362{#1}\fi}
\def\widetilde#1{\setbox\z@\hbox{$\m@th#1$}%
 \ifdim\wd\z@>\tw@ em\mathaccent"0\msbfam@5D{#1}%
 \else\mathaccent"0365{#1}\fi}
\font\teneufm=eufm10 scaled \@scale
\font\seveneufm=eufm7 scaled \@scale
\font\fiveeufm=eufm5  scaled \@scale
\newfam\eufmfam
\textfont\eufmfam=\teneufm
\scriptfont\eufmfam=\seveneufm
\scriptscriptfont\eufmfam=\fiveeufm

\csname amssym.def\endcsname

\begin{document}
\language=0
\begin{titlepage}
\hfill{QFT-TSU-17/96}\\
\hspace*{\fill}{hep-th/9702017}
\vspace*{3cm}
\begin{center}
\large{\bf N = 1, D = 3 Superanyons, $\bf osp(2|2)$ and \\
the Deformed Heisenberg Algebra}
\vspace{1.0cm}

\large{ I.V. Gorbunov\footnote{E-mail: ivan@phys.tsu.tomsk.su},
S.M. Kuzenko 
and S.L. Lyakhovich 
}\\

\footnotesize{{\it Department of Physics, Tomsk State University\\
Lenin Ave. 36, Tomsk, 634050 Russia} } \\
\end{center}
\vspace{1.5cm}

\begin{abstract}
We introduce $N=1$ supersymmetric generalization of the mechanical system
describing a particle with fractional spin in $D=1+2$ dimensions and being
classically equivalent to the formulation based on the Dirac monopole
two-form. The model introduced possesses hidden invariance under $N=2$
Poincar\'e supergroup with a central charge saturating the BPS bound.
At the classical level the model admits a Hamiltonian formulation with two
first class constraints on the phase space $T^\ast({\Bbb R}^{1,2})\times {\cal
L}^{1|1}$, where the K\"ahler supermanifold ${\cal L}^{1|1}\cong
OSp(2|2)/U(1|1)$ is a minimal superextension of the Lobachevsky plane. The
model is quantized by combining the geometric quantization on ${\cal
L}^{1|1}$ and the Dirac quantization with respect to the first class
constraints.  The constructed quantum theory describes a
supersymmetric doublet of fractional spin particles. The space of quantum
superparticle states with a fixed momentum is embedded into the Fock space of
a deformed harmonic oscillator.
\end{abstract}
\today
\end{titlepage}
\newpage
\section{Introduction}
Anyons \cite{Wil82}, particles with fractional spin and statistics
\cite{LeiMyr77,GoldMenSh8081} in $(1+2)$-dimensional space-time,
are not purely theoretical concept originating, for instance,
in the framework of field theory in the
presence of Chern-Simons field [4\,--\,7].
Several physical phenomena like the fractional Hall effect [8\,--\,10]
and the high-$T_c$ superconductivity \cite{Wilczekbook90} can be explained
on the base of this concept.

Last years there was a considerable interest in the study of
point-particle models of anyons [12\,--\,21],
mainly due to
possibility to derive a field theory for anyons by quantizing
a classical mechanical system in $D=1+2$ dimensions.
Up to now it is the most
successful approach to realize the quantum anyon states
by using the fields transforming in  unitary irreducible representations
of the universal covering
group of $\overline{SO^{\uparrow}(1,2)}\cong\overline{SU(1,1)}$
[12\,--\,15, 21\,--\,24].
These representations are infinite-dimensional
and, hence, an infinite set of
equations are required to single out one independent physical component.
Although various versions of such equations have been already suggested
(Refs.\ [12\,--\,15, 21\,--\,24]),
the problem remains open to realize them in a form appropriate
for accounting anyon self-interactions what is indispensable
for the construction of quantum field theory.

A convenient formulation of free field equations for fractional spin
particles was suggested in Ref.\ \cite{SorTkVol89SorVol93}.
In their approach, both the mass-shell constraint and the spin fixing
condition (which are imposed as independent equations in other models
[12\,--\,14, 16, 21])
originate as integrability conditions for the field equations of motion.
This was achieved by making use of the well known realization of
$so(1,2)$ as the Lie algebra of quadratic polynomials of the creation and
annihilation operators of the harmonic oscillator.  As a
consequence, only the particles with spins $(2n+1)/4, n=0,1,2\ldots$, (called
semions) appear in spectrum of the model \cite{SorTkVol89SorVol93}.
Recently, it has been recognized \cite{Ply96} that in order to extend
the semion construction \cite{SorTkVol89SorVol93} to the case of arbitrary
fractional spin particles one should make use of the deformed Heisenberg
algebra (DHA) (see \cite{Vas89,Ply96} and references therein) and
the superalgebra $osp(2|2)$. Thereby the one-particle anyon states can be
realized in the ${\Bbb Z}_2$-graded Fock space of the deformed quantum
oscillator, where the grading is induced by the Klein operator being
one of the generators of the DHA. These results imply that the DHA is
of primary importance for the description of anyon dynamics.

In the present paper we demonstrate that
the DHA naturally originates in the quantum supersymmetric theory of anyons.

We introduce $N=1$, $D=3$ super Poincar\'e invariant action for
a massive fractional spin superparticle living in
${\Bbb R}^{3|2}\times {\cal L}$, where ${\Bbb R}^{3|2}$  denotes the $N=1$,
$D=3$ flat superspace and ${\cal L}$ the Lobachevsky plane.  This mechanical
system is a minimal supersymmetric extension of special anyon model proposed
in \cite{GorKuzLyak}.  Our interest to the latter is caused by the fact that
the model proves to be classically equivalent to the formulation based on the
monopole-like symplectic two-form [17\,--\,21]
and, hence, allows introduction of coupling to arbitrary background fields.
On the other hand it can be treated as a reduction of the
$D=(1+3)$-dimensional massive spinning particle model developed in
\cite{LyakSegShar96}.

By construction, the model under consideration is manifestly $N=1$
supersymmetric. But it turns out to possess hidden invariance with respect to
$N=2$ Poincar\'e supergroup with a central charge saturating the
BPS bound (see, for instance, \cite{sohn}) on the mass shell.
As is well known, this the condition on central charge corresponds to
shortening of $N=2$ massive supermultiplets. The appearance of $N=2$
supersymmetry has a remarkable counterpart in Hamilton formulation of the
theory. Namely, the dynamics can be restricted on a surface of second class
constraints in such a way that it takes the form of the mechanics on the
phase space $T^\ast({\Bbb R}^{1,2})\times {\cal L}^{1|1}$, where the K\"ahler
supermanifold ${\cal L}^{1|1}=OSp(2|2)/U(1|1)$ (of complex dimension $1+1$)
presents itself a minimal superextension of the Lobachevsky plane.
$OSp(2|2)$ emerges as the group of all superholomorphic canonical
transformations on ${\cal L}^{1|1}$.

The $N=2$ Poincar\'e superalgebra with central charge and
the superalgebra $osp(2|2)$ prove to be closely related to each other at the
classical and quantum levels. Let us comment this crucial point in more
detail. In Hamilton approach the dynamics is governed by one first class
and six second class constraints. The second class constraints have a
complicate nonlinear structure what makes practically impossible a literal
application of the Dirac canonical quantization  (probably, it is the reason
why superanyon models have not been quantized until now). Our solution to the
problem is as follow.  We first reduce the dynamics with respect to four
second class constraints thereby arriving to the phase space $T^\ast({\Bbb
R}^{1,2})\times {\cal L}^{1|1}$. As a consequence, the superalgebra
$osp(2|2)$ is naturally realized in terms of the nonlinear Poisson bracket.
Special structure of the reduced phase space makes it possible to apply the
Berezin-Kostant quantization method
\cite{Berezin,Kost} for the inner
phase space, which has been recently developed for the supermanifold ${\cal
L}^{1|1}$ [32\,--\,34].
On $T^\ast({\Bbb
R}^{1,2})\times {\cal L}^{1|1}$, the rest constraints (one of first class and
two of second class) are equivalent to two first class constraints. In
quantum theory, the operatorial fulfillment of these constraints proves to be
equivalent to the requirement of $N=2$ Poincar\'e superalgebra to be
consistent quantum mechanically.  Thus, combining the geometric quantization
in ${\cal L}^{1|1}$ for the second class constraints and the Dirac
quantization with respect to the first class constraints, one can quantize
the superparticle with arbitrary (fixed) fractional superspin.  Short massive
representations of the $N=2$ Poincar\'e superalgebra with central charge are
realized on the fields transforming in atypical unitary representations of
$osp(2|2)$.  Moreover, the known connection between unitary representations
of $osp(2|2)$ and the DHA makes possible an alternative elegant realization
of the superanyon doublet in the Fock space of the deformed harmonic
oscillator.

The paper is organized as follows. The anyon model on the configuration
space ${\Bbb R}^{1,2}\times {\cal L}$ and its quantization are considered
in section \ref{s2}. In section \ref{s3} we analyze the $N=1$, $D=3$
superanyon model.  The global symmetries of the model and the structure of
the reduced phase space are studied in detail.  Section \ref{s4} is devoted
to the quantization of the superanyon model. Summary and concluding remarks
are given in section \ref{s5}.  In Appendix A we collect the conventions used
throughout the paper.  In Appendix B we describe the realization of the
Lobachevsky plane as a homogeneous space of the Lorentz group.

\section{Anyon model on ${\Bbb R}^{1,2}\times {\cal L}$}\label{s2}

As a starting point for supersymmetric extension, let us consider a model
of the fractional spin particle which was proposed in
\cite{GorKuzLyak}. The configuration space of the model
${\Bbb R}^{1,2}\times {\cal L}$, where ${\cal L}\cong SU(1,1)/U(1)$ denotes
a Lobachevsky plane,
is a homogeneous space of the $D=3$ Poincar\'e group.
The model is described by the following action functional
\begin{equation}
S=\int {\rm d}\tau\, L\qquad
L=m(\dot{x},n)+
i s \frac{\bar{z}\dot{z}-\dot{\bar{z}}z}{\zeta}\;,
\label{1}
\end{equation}
where
\begin{displaymath}
n_a\equiv
\frac{\zeta_a}{\zeta}
=-\left(\frac{1+z\bar{z}}{1-z\bar{z}}\;,\;
\frac{z+\bar{z}}{1-z\bar{z}}\;,\;i\frac{z-\bar{z}}{1-z\bar{z}}\right)\;
   \qquad n^2\equiv -1\;.
\end{displaymath}
Here $x^a$ and $z,\bar{z}$ are co-ordinates\footnote{The Lobachevsky space
${\cal L}$
is realized as the unit disc of complex plane, $|z|<1$.}
on ${\Bbb R}^{1,2}$ and $\cal L$ respectively,
$\zeta_a$ and $\zeta$ are defined by Eqs.\ (\ref{b4}) and (\ref{b7}),
$m$ and $s$ denote the mass and spin of the particle.
The model possesses global invariance with respect to the Poincar\'e group.
Infinitesimal Poincar\'e transformations
(with $f^a$ and $\omega^a$  parameters of translations and Lorentz
transformations) read
\begin{equation}
\renewcommand{\theequation}{\arabic{equation}.a}
\delta x^a=f^a \qquad \delta z=\delta{\bar z}=0 \label{3a}
\end{equation}
\addtocounter{equation}{-1}\renewcommand{\theequation}{\arabic{equation}.b}%
\begin{equation}
\delta x^a=\epsilon^{abc}x_b\;\omega_c \qquad \delta z=-i(\omega,\xi)\qquad
\delta{\bar z}=i(\omega,\bar{\xi}) \label{3b}\;,
\end{equation}
\setcounter{numerfive}{\value{equation}}%
\renewcommand{\theequation}{\arabic{equation}}%
where the vector-like objects $\xi_a\,,\bar{\xi_a}$ are defined
by Eq.\ (\ref{b5}). The Lagrangian (\ref{1}) is manifestly
translation-invariant, whereas the Lorentz transformations change it by total
derivatives of the form
\begin{equation}
\delta L=-\frac{s}{2}\frac{\rm d}{\rm d\tau}
( \frac{\partial }{\partial z}\xi_a
+\frac{\partial }{\partial \bar{z}}\bar{\xi}_a)\omega^a
\;.  \label{2}
\end{equation}
Really, by virtue of Eqs.\ (\ref{b2}), (\ref{b4}) and (\ref{b7}),
$n_a$ transforms as a three-vector, hence the first
term in the action functional is  manifestly Poincar\'e invariant. As to the
second term, it can be written as $s\int \Sigma_0$,
with the one-form $\Sigma_0$ being a solution of the equation
${\rm d}\Sigma_0=\Omega_0$, for the Lorentz invariant K\"ahler two-form
\begin{equation}
\Omega_0=-2i\frac{{\rm d}z\wedge {\rm d}\bar{z}}{\zeta^2}\label{b6}
\end{equation}
associated to the Lobachevsky plane.
The Lorentz invariance of $\Omega_0$ implies that $\Sigma_0$ may get
exact contributions under (\ref{3b}), and Eq.\ (\ref{2}) tells us this is
really the case.

The global symmetries related to the Poincar\'e group generate all the
independent Noether currents of the model. Here it is worth pointing out the
existence of another global space-time symmetry of the action functional
\begin{equation}
\delta x^a=- \varrho n^a \qquad \delta z=0\;, \label{3c}
\end{equation}
where $\varrho$ is a constant parameter. This rather unusual transformation
commutes with the Poincar\'e ones, and the associated Noether current is
trivial.  The point is that $\dot{z}=0$ on the equations of motion, hence
$n^a$ appears to be constant on the mass-shell.  Therefore Eq.\ (\ref{3c})
reduces to special space-time translations on-shell.

Since $L$ is a first-order homogeneous function of velocities,
the action remains invariant under world-line reparametrizations
of the form
\begin{equation}
\delta_{\epsilon}x^a=\dot{x}^a\epsilon (\tau)\qquad
\delta_{\epsilon} z =\dot{z}\epsilon (\tau)\qquad
\delta_{\epsilon}\bar{z}=\dot{\bar{z}}\epsilon (\tau)\;,  \label{7}
\end{equation}
where the parameter $\epsilon (\tau)$ being arbitrary modulo standard
boundary conditions.

Remarkable features of the model become transparent in the
Hamiltonian formalism. All the relations defining canonical momenta
conjugate to $x^a, z, \bar{z}$ constitute the set of primary constraints:
\begin{equation}
T_a=p_a-mn_a\approx 0\label{8}
\end{equation}
\begin{equation}
T=p_z-is\frac{\bar{z}}{\zeta}\approx 0\qquad
\overline{T}=p_{\bar{z}}+is\frac{z}{\zeta}\approx 0\;.\label{9}
\end{equation}
The Hamiltonian is a linear combination of these constraints.
There are no secondary constraints and Eqs.\ (\ref{8}), (\ref{9}) describe
the complete set of constraints in the model. The matrix of (canonical)
Poisson brackets of the constraints (\ref{8}), (\ref{9}) turns out to have
rank equal to four, it is the maximally possible value for antisymmetric
$5\times 5$ matrices. Hence, we have four second class constraints and one
first class constraint.

It is expedient for further consideration to reduce the
dynamics on the surface of the constraints (\ref{9}).
For $s\neq 0$ the corresponding Dirac brackets
are denoted by $\{\;,\;\}^{\ast}$ and have the form
\begin{equation}
\{x^a\,,\,p_b\}^{\ast}=\delta^{a}{}_{b} \qquad
\{z\,,\,\bar{z}\}^{\ast}=-\frac{i}{2s}(1-z\bar{z})^2 \;,
\label{her}
\end{equation}
the rest brackets between variables equal to zero.
The reduced phase space obtained in this way is seen to be isomorphic
to the product of two symplectic manifolds,
$T^{\ast}({\Bbb R}^{1,2})\times {\cal L}$, where ${\cal L}$ catches
a standard nonlinear symplectic structure of the Lobachevsky plane
\cite{Berezin,Perelombook87}.

Let us discuss the physical content of the model.
First, consider the Hamiltonian generators of the Poincar\'e transformations
(\thenumerfive). For the energy-momentum vector ${\cal P}_a$ and
 the angle momentum vector ${\cal J}_a$ , one gets
\begin{equation}
{\cal P}_a=p_a\qquad {\cal J}_a=\epsilon_{abc}x^b p^c+ J_a\;,\label{4}
\end{equation}
where $J_a$ denotes the spin momentum vector
\begin{equation}
J_a= i\xi_a p_z -\frac{s}{2}\partial\xi_a-
i\bar{\xi}_a p_{\bar{z}}-\frac{s}{2}\bar{\partial}\bar{\xi}_a
=-sn_a \;.
\label{5}
\end{equation}
Here we have accounted the constraints (\ref{9}).
With respect to Poisson bracket (\ref{her}), the functions (\ref{4}) generate
the Poincar\'e algebra $iso(1,2)$, whereas
the spin generators (\ref{5}) span internal Lorentz algebra
$so(1,2)$  related to the automorphism group of the Lobachevsky plane.
Associated to the Poincar\'e generators (\ref{4}) are
the phase-space Casimir functions
${\cal P}^a{\cal P}_a=p^2$ and ${\cal P}^a{\cal J}_a=-s(p,n)$.
As a consequence of the  constraints (\ref{8}), they are identically conserved
\begin{equation}
\renewcommand{\theequation}{\arabic{equation}.a}
T^{(1)}=p^2+m^2\approx 0\label{11}
\end{equation}
\addtocounter{equation}{-1}\renewcommand{\theequation}{\arabic{equation}.b}%
\begin{equation}
T^{(2)}=(p,n)+m\approx 0\;. \label{12}
\end{equation}
\setcounter{numertwo}{\value{equation}}%
\renewcommand{\theequation}{\arabic{equation}}%
One can also verify that functions of
the Poincar\'e generators exhaust all physical observables in the
model\footnote{Physical observables are understood as phase space functions
commuting (with respect to the Poisson bracket (\ref{her}))
with the first class
constraints (\thenumertwo).}. Therefore the model describes the
irreducible dynamics of $D=3$ particle with mass $m$ and spin $s$. Besides
the particle energy $p^0$ is positive, as a consequence of Eq.\ (\ref{8}).

Remarkably, the mixed first and second class constraints (\ref{8})
proves to be equivalent to the first class constraints
(\thenumertwo). This immediately follows
from the decomposition
\begin{equation}
p_a\equiv2\frac{(p,\xi)}{\zeta^2}\bar{\xi_a}+
2\frac{(p,\bar{\xi})}{\zeta^2}\xi_a-(p,n)n_a\,,
\label{13}
\end{equation}
which is true for arbitrary three-vector $p_a$, in virtue of Eq.\ (\ref{B8}).
Really, the constraints (\ref{8}) imply $(p,\xi)=(p,\bar{\xi})=0$,
hence Eqs.\
(\thenumertwo) are fulfilled. On the other hand, by squaring Eq.\
(\ref{13}) one gets \begin{equation}
4\left|\frac{(p,\xi)}{\zeta}\right|^2\equiv p^2+(p,n)^2\;.\label{14}
\end{equation} Thus, the constraints (\thenumertwo) imply $(p,\xi
)=(p,\bar{\xi})=0$.  Hence, the set of three constraints (\ref{8}) (among
which there are two second class ones and one of first class) are equivalent
to the pair of first class constraint (\thenumertwo).  The above
observation will be crucial for quantization.

On the mass shell (\ref{11}),
Eq.\ (\ref{8}) can be treated as a parametrization of the
mass hyperboloid by local complex co-ordinates $z\,,\,\bar{z}$.
This means, however, that we can rewrite the two-form (\ref{b6}) in the way
\begin{equation}
\Omega_0=\frac{1}{2}\frac{\epsilon^{abc}p_a {\rm d}p_b\land
{\rm d}p_c}{(-p^2)^{3/2}}\;,\label{15}
\end{equation}
that is as a Dirac
monopole two-form. Consequently, our model proves to be a reformulation of
the well known anyon models based on the monopole-like two-form
[17\,--\,21].
This fact can
be alternatively established by deriving the Dirac brackets (to be
denoted below by $\{\;,\;\}^{\ast\ast}$) associated to the second class
constraints
${(p,\xi )=0}$, ${(p,\bar{\xi})=0}$.  These brackets have the explicit
structure
\begin{equation}
\{x^a\,,\,x^b\}^{\ast\ast}=s\frac{\epsilon^{abc}p_c}{(-p^2)^{3/2}}\qquad
\{x^a\,,\,p_b\}^{\ast\ast}=\delta^{a}{}_{b}\qquad
\{p_a\,,\,p_b\}^{\ast\ast}=0 \label{16.b}
\end{equation}
\setcounter{numerthree}{\value{equation}}%
and reproduce the Poisson brackets for the particle models mentioned.

In a sense the ``minimal'' approach based on the two-form (\ref{15}) and
Poisson bracket (\thenumerthree) appears to be very natural. In particular,
there is no problem to introduce consistent coupling to external fields
[17\,--\,20].
However, the realization of quantization scheme
in terms of ``nonlocalizable'' co-ordinates
has become a difficult problem in view of the complicated structure of
the Poisson brackets for co-ordinates.
Moreover, it is not possible in this approach to introduce
``localizable'' co-ordinates without loss of manifest
covariance~\cite{CorPly95}.
To the contrary, the formulation on the extended phase space
$T^{\ast}({\Bbb R}^{1,2})\times {\cal L}$ admits a natural
quantization scheme we are going to describe.

To quantize the model, we shall make use of the following prominent
features of the model. First, all physical observables, which are
phase space functions commuting with the first class constraints, are
actually functions of the Poincar\'e generators (\ref{4}) only.
Thus the quantization problem is to construct an  appropriate realization for
the unitary representations of the Poincar\'e group.  Classically,
the Poincar\'e generators (\ref{4}) in the phase space of the model are
splitted into two pieces, one of which includes only space-time variables and
another corresponds to the internal space ${\cal L}$. It is the latter part
of the generators  which is relevant for nontrivial spin values.  Second, one
can observe that the spin part (\ref{5}) of ${\cal J}_a$  coincides with the
covariant Berezin symbols of the group $\overline{SU(1,1)}$ on the
Lobachevsky plane \cite{Berezin,Perelombook87}.  In view of all the features
mentioned, it seems sensible to combine the Dirac canonical quantization for
the Minkowski degrees of freedom with a geometric quantization for spin.

We realize the Hilbert space of
one-particle anyon states of mass $m$ and spin ${s>0}$
as a space of functions $F(p,\bar{z})$ ,
$F:{\Bbb R}^{1,2}\times {\cal L} \rightarrow {\Bbb C}$
to be antiholomorphic\footnote{This particular realization is useful, since
it provides the correspondence principle for Eqs.\  (\ref{her}), (\ref{4}) and
(\ref{5}) and gives the proper energy sign.} on the Lobachevsky plane (that
is, antiholomorphic in the unit disk of $\Bbb C$, $|z|<1$). The operator
realization of the classical Poincar\'e generators ${\cal P}_a$ and ${\cal
J}_a$ (\ref{4}) reads
\begin{equation}
\begin{array}{c}\displaystyle
\hat{{\cal P}}_a=p_a\qquad\hat{{\cal
J}}_a=-i\epsilon_{abc}p^b\frac{\partial}{\partial
p_c}+\hat{J}^s_a\;,\\ \displaystyle
\hat{J}^s_a=-\bar{\xi}_a\bar{\partial}-s\bar{\partial}\bar{\xi_a},
\label{17}
\end{array}
\end{equation}
where
$\bar{\partial}\equiv\partial/\partial \bar{z}$.
The generators (\ref{17}) are Hermitian with respect to the following inner
product
\begin{equation}
\langle F|G\rangle =(2s-1)\int {\rm d}^3 p\int\nolimits_{{\cal L}}
\frac{{\rm d}z{\rm d}\bar{z}}{2\pi
i}\zeta^{2s-2}\overline{F(p,\bar{z})}G(p,\bar{z})\;. \label{18}
\end{equation}

To complete the quantization, we impose operator counterparts of
the first-class constraints (\thenumertwo) on the physical states
$F^{phys}$:
\begin{equation}
\begin{array}{l}
(p^2+m^2)F^{phys}(p,\bar{z})=0\\ \label{19}
\!\!\left( (p,\hat{J}^s)-ms\right) F^{phys}(p,\bar{z})=0\;.
\end{array}
\end{equation}
Our construction corresponds to the well known realization
(see e.g. \cite{Ply911,JacNair91,CorPly95}) of the
$D=3$ Poincar\'e group representations of mass $m$ and spin $s>0$ in
terms of
infinite-component fields transforming by an appropriate irreducible
unitary representation of discrete series of $D^{s}_{+}$ of the group
$\overline{SU(1,1)}$ bounded below [35\,--\,37, 23].
The components $F_n(x),\,n=0,1,2,\ldots$ of the fields are obtained by the
series expansion of our wave functions in
$|n\rangle\equiv(\Gamma(2s+n)/\Gamma(n+1)\Gamma(2s))^{1/2}\bar{z}^n$. Finally
let us note that the case of $s<0$ can be treated in a similar way
by the use of the representation of $\overline{SU(1,1)}$ belonging to the
other discrete series $D^{-s}_{-}$ bounded above.

\section{Superparticle dynamics on
$T^\ast({\Bbb R}^{1,2})\times{\cal L}^{1|1}$ } \label{s3}

The simplest way to obtain a supersymmetric generalization of the model
described is to extend the configuration space to a supermanifold
${\Bbb R}^{3|2}\times {\cal L}$, where the Grassmann sector is parametrized
by an anticommuting Majorana spinor
$\theta^{\alpha}$,\footnote{The reality conditions on spinors
in the $SU(1,1)$ formalism are described in Appendix A.} and
to substitute $\dot{x}^a$ in the action by
$\Pi^a=\dot{x}^a-i(\sigma^a)_{\alpha\beta}\theta^\alpha\dot{\theta}^\beta\;.$
Then, one results with the $N=1$, $D=3$ superanyon theory\footnote{The case
of extended supersymmetry, $N>1$, deserve special treatment and will be
considered elsewhere.} with the action
functional
\begin{equation}
S=\int {\rm d}\tau\, L\qquad
L=m(\Pi,n)+
i s \frac{\bar{z}\dot{z}-\dot{\bar{z}}z}{\zeta}\qquad \label{20}
\end{equation}
By construction, the model possesses global symmetry with respect to
the $N=1$ Poincar\'e
supergroup, and the corresponding infinitesimal transformations read
\begin{equation}
\begin{array}{lll}\displaystyle
\delta x^a=f^a &\displaystyle \delta z=0&\displaystyle \delta\theta^\alpha=0
\\[1mm]\displaystyle
\delta{x^a}=i\epsilon^{\alpha}(\sigma^a)_{\alpha\beta}\theta^\beta&
\displaystyle
\delta z=0 & \displaystyle \delta\theta^\alpha=\epsilon^\alpha\\
\displaystyle \delta x^a=\epsilon^{abc}x_b\;\omega_c
&\displaystyle \delta z=-i(\omega,\xi) &\displaystyle
\delta\theta^\alpha=\frac{i}{2}\omega^\alpha{}_\beta\theta^\beta\;.
\label{20a}
\end{array}
\end{equation}
Here $f^a$, $\omega^a$ and $\epsilon^\alpha$ are the parameters of
translations, Lorentz and supersymmetry transformations respectively.
Similarly to the non supersymmetric model (\ref{1}), Lorentz transformations
change the Lagrangian (\ref{20}) by total derivatives.

Along with the dynamical symmetries (\ref{20a}), the theory
possesses several invariances which do not lead to new independent Noether
currents. Such global symmetries are described by
the following transformations
\begin{equation}
\begin{array}{lll}
\delta x^a=-\varrho n^a& \delta z=0 &\qquad\delta\theta^\alpha=0\\
\delta x^a=0 & \delta z=0&\qquad
\delta\theta^\alpha=-2i\mu n^\alpha{}_\beta\theta^\beta\\
\delta{x^a}=-\eta^{\alpha}(\sigma^a)_{\alpha\beta}n^\beta{}_\gamma
\theta^\gamma & \delta z=0
&\qquad\delta\theta^\alpha=-in^\alpha{}_\beta\eta^\beta\;,
\end{array} \label{21a}
\end{equation}
where $\varrho$ and $\mu$ are bosonic infinitesimal parameters and
$\eta^\alpha$ Grassmann ones,
$n^\alpha{}_\beta\equiv(n^a\sigma_a)^\alpha{}_\beta$ is constructed in terms
of $z,\bar z$ like as in Eq.\ (\ref{1}). The transformations (\ref{20a}) and
(\ref{21a}) turn out to generate a closed superalgebra off the mass-shell.
To analyze the structure of that superalgebra,
it is convenient to pass to the Hamiltonian formalism.

Introducing the momenta conjugate to $x^a,z,\bar z,\theta^\alpha$ and
defining the canonical gra\-ded Poisson brackets \begin{displaymath}
\{x^a\;,\;p_b\}=\delta^a{}_b\qquad \{z\;,\;p_z\}=\{\bar{z}\;,\;p_{\bar{z}}\}=1
\qquad \{\theta^\alpha\;,\;\pi_\beta\}=\delta^\alpha{}_\beta\;,
\end{displaymath}
we observe that the model contains the following set of constraints
\begin{equation}
T_a=p_a-mn_a\approx 0\label{25}
\end{equation}
\renewcommand{\theequation}{\arabic{equation}.a}
\vspace{-5mm}
\begin{equation}
T_\alpha=\pi_\alpha+imn_{\alpha\beta}\theta^\beta
\approx 0\label{26.a}
\end{equation}
\addtocounter{equation}{-1}
\renewcommand{\theequation}{\arabic{equation}.b}
\vspace{-5mm}
\begin{equation}
T=p_z-is\frac{\bar{z}}{\zeta}\approx 0\qquad
\overline{T}=p_{\bar{z}}+is\frac{z}{\zeta}\approx 0\label{26.b}
\end{equation}
\renewcommand{\theequation}{\arabic{equation}}%
\setcounter{numerone}{\value{equation}}%
which involve six constraints of the second class and one of the
first class. As it is obvious, the first class constraint generates
world-line reparametrizations and thus the physical Hamiltonian is zero.
The Hamiltonian generators of the super Poincar\'e transformations (\ref{20a})
look like
\begin{equation}
{\cal P}_a=p_a \qquad {\cal J}_a=\epsilon_{abc}x^b p^c+ J_a \qquad
{\cal Q}^{1}_\alpha=i p_{\alpha\beta}\theta^\beta-\pi_\alpha\;,
\label{21}
\end{equation}
where
\begin{equation}
J_a=-\frac{i}{2}(\sigma_a)_{\alpha\beta}\theta^\alpha\pi^\beta+i\xi_a
p_z -\frac{s}{2}\partial\xi_a- i\bar{\xi}_a p_{\bar{z}}
-\frac{s}{2}\bar{\partial}\bar{\xi}_a\;.
\label{22}
\end{equation}
Further, the generators of transformations (\ref{21a}) have the form
\begin{equation}
{\cal Z}=-(p,n) \qquad {\cal K}=2in_{\alpha\beta}\theta^\alpha\pi^\beta
\qquad
{\cal Q}^{2}_\alpha=-p_{\alpha\beta}n^\beta{}_\gamma\theta^\gamma
+in_\alpha{}^\beta\pi_\beta\;.\label{23}
\end{equation}
The generators (\ref{21}) and (\ref{23}) prove to satisfy the (anti)
commutation relations
\begin{eqnarray}
&&\{{\cal J}_a\;,\;{\cal J}_b\}=\epsilon_{abc}{\cal J}^c \qquad \qquad \quad
\{{\cal J}_a\;,\;{\cal P}_b\}=\epsilon_{abc}{\cal P}^c \nonumber \\
&&\{{\cal J}_a\;,\;{\cal Q}^I_\alpha\}=\frac{i}{2}(\sigma_a)_{\alpha\beta}
{\cal Q}^{I\;\beta} \qquad
\{{\cal Q}^I_\alpha\;,\;{\cal K}\}=2\epsilon^{IJ}{\cal Q}^J_\alpha
\label{24} \\
&&\{{\cal Q}^I_\alpha\;,\;{\cal Q}^J_\beta\}=
-2i\delta^{IJ}p_{\alpha\beta}-2\epsilon^{IJ}
\epsilon_{\alpha\beta}{\cal Z}\;, \nonumber
\end{eqnarray}
the rest brackets being equal to zero, where
$I,J=1,2$, $\epsilon^{IJ}=-\epsilon^{JI}$, $\epsilon^{01}=1$.
What we have obtained is $N=2$ Poincar\'e superalgebra with a central charge
described by $\cal Z$ and $U(1)$ isotopic charge $\cal K$ acting on the
internal index of ${\cal Q}^I_\alpha$.
The functions
(\ref{21}) generate $N=1$ subalgebra.

Let us discuss in more detail the system of constraints
(\ref{25}) and (\thenumerone) which are different from that
defined by Eqs.\ (\ref{8}) and (\ref{9}) by the presence of fermionic
constraints (\ref{26.a}). The latter can be rewritten in a more familiar,
for superparticle models, form
\begin{equation}
T^{\prime}_\alpha=\pi_\alpha+ip_{\alpha\beta}\theta^\beta\approx 0
\label{26*}
\end{equation}
on the surface of constraints (\ref{25}).
We prefer, however, to use the original representation (\ref{26.a})
in which the fermionic constraints
do not involve the space-time variables and admit an interesting geometric
interpretation related to the reduction (for $s>0$, $m>0$) on the surface of
second class constraints (\thenumerone).  To explain this interpretation,
write down the respective Dirac brackets:
\begin{eqnarray}
\{z\,,\,\bar{z}\}^{\ast}= -\frac{i\zeta^2}{2s}\left(1+\frac{1}{2}
\frac{\theta\bar{\theta}}{\zeta}\right) & &
\{\theta^\alpha\,,\,\theta^\beta\}^{\ast}=
-n^{\alpha\beta}\frac{i}{2m}\left(1-\frac{1}{2}
\frac{\theta\bar{\theta}}{\zeta}\right) \nonumber\\
\{z\,,\,\theta^\alpha\,\}^{\ast}=\frac{i}{2 \sqrt{2ms}}z^\alpha\theta & &
\{\bar{z}\,,\,\theta^\alpha\}^{\ast}=\frac{i}{2 \sqrt{2ms}}\bar{z}^\alpha
\bar{\theta}\;, \label{27}
\end{eqnarray}
and the rest Dirac brackets involving the space-time variables keep
their canonical form, that is they vanish except
$\{x^a\,,\,p_b\}^{\ast}=\delta^{a}{}_{b}$.
Here
\begin{equation}
\sqrt{\frac{s}{2m}}\theta \equiv z_\alpha\theta^\alpha = z\theta^0-\theta^1
\qquad
\sqrt{\frac{s}{2m}}\bar{\theta} \equiv \bar{z}_\alpha\theta^\alpha =
\theta^0-\bar{z}\theta^1
\label{29}
\end{equation}
and the twistor-like variables $z^\alpha,\bar{z}^\alpha$ are defined in
Appendix B. The complex Grassmann variable $\theta$ is
in a one-to-one correspondence
with Majorana spinor $\theta^\alpha$ and, together with its complex
conjugate $\bar\theta$, can be used to parametrize the odd sector of the
constrained surface.  From (\ref{27}) one deduces
\begin{equation}
\{\theta\,,\,\bar{\theta}\}^{\ast}= \frac{i}{s}\zeta\left(1+\frac{1}{2}
\frac{z\bar{z}\theta\bar{\theta}}{\zeta}\right)\qquad
\{z\,,\,\bar{\theta}\}^{\ast}=\frac{i\zeta}{2s}z\bar{\theta}\qquad
\{\bar{z}\,,\,\theta\}^{\ast}=-\frac{i\zeta}{2s}\bar{z}\theta\,.
\label{30}
\end{equation}
Eqs.\ (\ref{27}) and (\ref{30}) mean that the symplectic
structure on the reduced phase space is induced by the two-superform
\begin{displaymath}
\Lambda = {\rm d}p_a\wedge {\rm d}x^a + s\Omega ,
\end{displaymath}
where
\begin{equation}
\Omega=2i\left(1-\frac{1}{2}\frac{1+z\bar{z}}{\zeta}
\theta\bar{\theta}\right)\frac{{\rm d}z\wedge {\rm d}\bar{z}}{\zeta^2}
+i\left(\frac{{\rm d}\theta \wedge {\rm d}\bar{\theta}}{\zeta}
-\frac{\bar{z}\theta}{\zeta^2}{\rm d}z \wedge {\rm d}\bar{\theta}
-\frac{z\bar{\theta}}{\zeta^2}{\rm d}\bar{z} \wedge {\rm d}\theta\right).
\label{31}
\end{equation}
We follow Berezin's conventions for superforms \cite{Berezin}
(see Appendix A). It is easy to note that $\Omega$ can be represented as
follows
\begin{equation}
\Omega=i( {\rm d}z\frac{\partial}{\partial z}
+{\rm d}\theta\frac{ {\stackrel{\rightarrow}{\partial}} }{\partial\theta})
\wedge ( {\rm d}\bar{z}\frac{\partial}{\partial\bar{z}}
+{\rm d}\bar{\theta}
\frac{ {\stackrel{\rightarrow}{\partial}} }{\partial\bar{\theta}})\Phi \;,
\label{mud}
\end{equation}
where
\begin{equation}
\Phi(z,\bar{z},\theta,\bar\theta)= -2\ln{(\zeta +\frac{1}{2}\theta\bar{\theta}
)}=-2\ln{\zeta}-\frac{\theta\bar{\theta}}{\zeta}\,.
\label{32}
\end{equation}
We conclude that $\Omega$ and, hence, $\Lambda$ are closed,
${\rm d}\Lambda = {\rm d}\Omega=0$.

The above consideration shows that the reduced phase space
has the structure
of direct product of symplectic spaces
$T^\ast({\Bbb R}^{1,2})\times {\cal L}^{1|1}$, with
${\cal L}^{1|1} = {\cal L} \times {\Bbb C}{}^{0|1}$ being a complex
supermanifold (of dimension $1+1$) parametrized by the complex even $z$
and odd $\theta$ co-ordinates. The symplectic structure on
${\cal L}^{1|1}$ is determined by the closed non-degenerate
superform $\Omega$ which is in
fact a K\"ahler superform, in accordance with Eq.\ (\ref{mud}), and the
corresponding superpotential reads as in Eq.\ (\ref{32}).
This K\"ahler supermanifold has been introduced in Refs.
\cite{Grad93,GradNiet96} as coadjoint orbit of simplest orthosymplectic
supergroups (degenerate orbit of $OSp(2|2)$ and a regular
orbit of $OSp(1|2)$) and termed {\it superunit disk}. Therefore, ${\cal
L}^{1|1}$ is a homogeneous space \cite{GradNiet96} of the supergroup
$OSp(2|2)$, ${\cal L}^{1|1}=OSp(2|2)/U(1|1)$ (hence, it can also be realized
in the manner ${\cal L}^{1|1}=OSp(1|2)/U(1)$).

$OSp(2|2)$ turns out to be the group
of all {\it canonical\/} (with respect to $\Omega$)
{\it superholomorphic transformations\/} on ${\cal L}^{1|1}$.
Infinitesimally, these transformations look like
\begin{displaymath}
\delta z=-i\omega^a \xi_a-\epsilon_\alpha z^\alpha\theta\qquad
\delta\theta=-\frac{i}{2}\omega^a\frac{\partial}{\partial z}\xi_a \theta
-i\mu\theta+2\bar\epsilon_\alpha z^\alpha\;,
\end{displaymath}
where $\omega^a$, $\mu$ are bosonic real parameters and $\epsilon_\alpha$
fermionic complex ones.
The functions
\begin{equation}
\begin{array}{ll}\displaystyle
J_a=-sn_a\left(1-\frac{1}{2}\frac{\theta\bar{\theta}}{\zeta}\right)\quad&
\displaystyle\quad B=-s\left(1+\frac{1}{2}\frac{\theta\bar{\theta}}{\zeta}
\right)\\
\displaystyle
\theta^\alpha=\sqrt{\frac{s}{2m}}
\frac{z^\alpha\bar{\theta}-\bar{z}^\alpha\theta}{\zeta}\quad&
\displaystyle\quad
\pi_\alpha=i\sqrt{\frac{ms}{2}}\frac{z_\alpha\bar{\theta}+
\bar{z}_\alpha\theta}{\zeta}
\end{array}
\label{33}
\end{equation}
serve as the corresponding (real) generators of $OSp(2|2)$, and their algebra,
with respect to the Dirac bracket, reads
\begin{equation}
\begin{array}{lll} \displaystyle
\{J_a\;,\;J_b\}^\ast=\epsilon_{abc}J^c &
\displaystyle\{J_a\;,\;\theta^\alpha\}^\ast=\frac{i}{2}(\sigma_a)^{\alpha}
{}_{\beta}\theta^\beta & \displaystyle
\{J_a\;,\;\pi_\alpha\}^\ast=-\frac{i}{2}(\sigma_a)_{\alpha}
{}^{\beta}\pi_\beta \\[2mm]
\displaystyle\{\theta^\alpha\;,\;B\}^\ast=\frac{1}{2m}\pi^\alpha &
\displaystyle\{\pi_\alpha\;,\;B\}^\ast=-\frac{m}{2}\theta_\alpha &
\displaystyle\{J_a\;,\,B\}^\ast=0 \\[2mm]
\displaystyle\{\theta^\alpha\;,\;\theta^\beta\}^\ast=
\frac{i}{2ms} J^{\alpha\beta} & \displaystyle
\{\pi_\alpha\;,\;\pi_\beta\}^\ast=
\frac{im}{2s} J_{\alpha\beta} & \displaystyle
\{\theta^\alpha\;,\;\pi_\beta\}^\ast=
-\frac{1}{2s}\delta^{\alpha}{}_{\beta}B\,.
\end{array}
\label{34}
\end{equation}
The generators $J_a$
and $\theta^\alpha$ (or $\pi_\alpha$) form a superalgebra $osp(1|2)$.

Let us note that the role of $OSp(2|2)$ for the
superparticle model (\ref{20}) is similar to the internal Lorentz group
$SU(1,1)/{\Bbb Z}_2$, whose action is defined on ${\cal L}$ only,
in the particle
model of Sec.\ \ref{s2}. Really, in accordance with Eqs.\
(\ref{27}--\ref{31}) the reduced
phase space (the surface of constraints (\thenumerone)) of the superparticle
is isomorphic to $T^\ast({\Bbb R}^{1,2})\times {\cal L}^{1|1}$, whereas
its particle counterpart is $T^\ast({\Bbb R}^{1,2})\times {\cal L}$. $OSp(2|2)$
(respectively $SU(1,1)/{\Bbb Z}_2$) leaves invariant  the K\"ahler
two-superform $\Omega$ (\ref{31}) on ${\cal L}^{1|1}$ (respectively, the
K\"ahler two-form $\Omega_0$ (\ref{b6}) on ${\cal L}$). We introduce the
one-form $\Sigma_0$, ${\rm d}\Sigma_0=\Omega_0$, into the action functional
(the second term in (\ref{1})). $\Sigma_0$ changes
at most by total derivatives under the $SU(1,1)/{\Bbb Z}_2$ transformations.
Let us now rewrite the action functional (\ref{20}) in the form
\begin{displaymath}
S=\int \left(m n_a {\rm d}x^a-i\left[
m n_{\alpha\beta}\theta^\alpha {\rm d}\theta^\beta-
s\frac{\bar{z}{\rm d}z-z{\rm d}\bar{z}}{\zeta}
\right]\right)\;.
\end{displaymath}
It is easy to verify that the term in the square brackets is related to a
one-superform $\Sigma$ such that ${\rm d}\Sigma=\Omega$. Thus, $\Sigma$
changes at most by exact contributions under the $OSp(2|2)$ transformations.

It should be emphasized that neither $OSp(2|2)$ nor its non supersymmetric
analogue $SU(1,1)/{\Bbb Z}_2$ (the internal Lorentz group)
do not originate as symmetry (super) groups of the corresponding
mechanical systems. The true symmetry (super) groups of the models
(\ref{1}) and (\ref{20}) are the Poincar\'e group and its $N=1$
superextension respectively, which exhaust all global invariance
transformations giving rise to independent Noether currents. However, the
internal Lorentz algebra $so(1,2)$ and its superextension $osp(2|2)$
naturally appear in the Hamilton approach as building blocks of the (super)
Poincar\'e generators.  Really, we have seen that the Poincar\'e generators~%
(\ref{4}) in $T^\ast({\Bbb R}^{1,2})\times {\cal L}$ consist of two sectors,
one of which is associated with the space-time co-ordinates and momenta and
the second coincides with the $so(1,2)$ generators~(\ref{5}). A similar
phenomenon takes place in the superparticle model.  It is apparent
that on the constrained surface (\thenumerone) the generators of the
Poincar\'e supergroup become phase-space functions depending on $x^a$, $p_a$
and $OSp(2|2)$ generators~(\ref{33}). This observation will be of primary
importance when quantizing the model in the following section.

In spite of the strong analogy mentioned between the particle and
superparticle models,
there is an essential difference in realization of the global
symmetry groups in the
reduced phase spaces. The action of the Poincar\'e group is obviously
well defined on
$T^\ast({\Bbb R}^{1,2})\times {\cal L}$. At the same time,
supersymmetry  can not be globally realized
on $T^\ast({\Bbb R}^{1,2})\times {\cal L}^{1|1}$ and restores only
on the surface of the rest constraints~(\ref{25}). Straightforward
calculations of (anti) commutation relations of the generators~(\ref{21},~%
\ref{23}), with respect to the Dirac brackets, show that all the brackets
(\ref{24}) remain intact in the strong sense except $\{{\cal
Q}^I_\alpha\;,\;{\cal Q}^J_\beta\}^\ast$ and $\{{\cal Z}\;,\;{\cal
Q}^I_\alpha\}^\ast$. The latter can be presented in the manner
\begin{eqnarray} &&
\{{\cal Q}^I_\alpha\;,\;{\cal Q}^J_\beta\}^\ast=
-2i\delta^{IJ}p_{\alpha\beta}-2\epsilon^{IJ}\epsilon_{\alpha\beta}
{\cal Z} +(p^2+m^2)c^{(1)}{}^{IJ}_{\alpha\beta}
+((p,n)+m)c^{(2)}{}^{IJ}_{\alpha\beta}\nonumber \\&&
\{{\cal Z}\;,\;{\cal Q}^I_\alpha\}^\ast=(p^2+m^2)c^{(1)}{}^I_\alpha+
((p,n)+m)c^{(2)}{}^I_\alpha\;,\nonumber
\end{eqnarray}
where $c^{(\cdot)}{}^{IJ}_{\alpha\beta}$, $c^{(\cdot)}{}^I_\alpha$ are some
functions on $T^\ast({\Bbb R}^{1,2})\times {\cal L}^{1|1}$, whose explicit
expressions are rather cumbersome and not important here.
Hence the Poincar\'e superalgebra restores only on the surface
of constraints (\ref{25}).
Let us discuss this point in more detail.

Similarly to the constraints structure in the anyon model of Sec.\ 2\,,
Eq.\ (\ref{25}) describes two second class and one first class constraints
which are equivalent to
the pair of first class constraints (\thenumertwo). The latter can
be used to
evaluate the Casimir functions $C_1={\cal P}^a{\cal P}_a$ and
$C_2={\cal P}^a{\cal J}_a+\frac{1}{8}{\cal Q}^{I\,\alpha}{\cal Q}^I_\alpha
-\frac{1}{4}{\cal ZK}$
of $N=2$ Poincar\'e superalgebra, which turn out to conserve identically
on the total constraint surface.
Then we find that the model describes a superparticle with mass $m$,
superspin $s$, central charge ${\cal Z}=m$ and positive energy $p^0>0$.

Relation ${\cal Z}=m$ corresponds to saturating the BPS bound
$m \geq |{\cal Z}|$ for massive multiplets in extended supersymmetry.
The specific feature of such a choice is multiplet-shortening
through central charges \cite{sohn}. This is the case $m = |{\cal Z}|$
when a massive supermultiplet contains the same number of particles
as a massless one. Such massive multiplets are called hypermultiplets
\cite{sohn}.  In the case of $N=2$, $D=3$ Poincar\'e superalgebra,
a massive multiplet (superparticle) of superspin $s$
describe  a quartet of particles with
spins ${(s,s+\frac{1}{2},s+\frac{1}{2},s+1)}$ for $m > |{\cal Z}| $
and a doublet $(s,s+\frac{1}{2})$ for $ m= |{\cal Z}| $.
We conclude that our model describes a massive $N=2$ hypermultiplet of
superspin $s$ or, in other words, a supersymmetric doublet of anyons
with spins $s$ and $s +\frac 12$.

Because of the relation ${\cal Z}=m$,
not all Hamiltonian generators
(\ref{21}) and (\ref{23}) of the $N=2$ Poincar\'e superalgebra are
functionally independent, when restricted to
the total constraint surface (\ref{25}, \thenumerone), but only their
$N=1$ subset (\ref{21}). The rest generators can be expressed as follows
\begin{equation}
{\cal Q}^2_\alpha=-\frac{i}{m}p_\alpha{}^\beta{\cal Q}^1_\beta
\qquad {\cal K}=-\frac{1}{2m}{\cal Q}^{1\,\alpha}{\cal Q}^{1}_\alpha\qquad
{\cal Z}=-\frac{p^2}{m}=m  \label{ident}
\end{equation}
on the full constraint surface. Moreover, any physical
observable proves to be a function of the $N=1$ super Poincar\'e
generators (\ref{21}) only.

Eq.\ (\ref{ident}) shows that the hidden $N=2$
supersymmetry (\ref{21a}) can be treated as an artifact of the embedding of
$N=2$ Poincar\'e superalgebra into the universal enveloping algebra of $N=1$
one. The transformations (\ref{21a}) present themselves special linear
combinations of the $N=1$ transformations (\ref{20a}) with the coefficients
depending on the on-shell conserved quantities.

Concluding this section we consider the reduction to the surface of the rest
second class constraints $(p,\xi )=0\,,\,(p,\bar{\xi})=0$.
The reduced phase space is originated from the symplectic two-superform
\begin{displaymath}
\Lambda={\rm d}p_a\wedge {\rm d}x^a + s\Omega
\vspace{-0.1cm}
\end{displaymath}
\begin{equation}
\Omega= \frac{1}{2}\frac{\epsilon^{abc}p_a {\rm d}p_b\land {\rm
d}p_c}{(-p^2)^{3/2}}
+\frac{im}{s\sqrt{-p^2}}(\eta_{ab}-
\frac{p_ap_b}{p^2})\theta^\alpha(\sigma^a)_{\alpha\beta}{\rm d}p^b \wedge
{\rm d}\theta^\beta-\frac{im}{s\sqrt{-p^2}}p_{\alpha\beta}{\rm d}\theta^\alpha
\wedge {\rm d}\theta^\beta
\label{kok}
\end{equation}
The respective nonvanishing Dirac brackets are
\begin{equation}
\begin{array}{ll}\displaystyle
\{x^a\,,\,x^b\}^{\ast\ast}=s\frac{\epsilon^{abc}p_c}{(-p^2)^{3/2}}
\left(1-\frac{m}{2s}\theta^\alpha\theta_\alpha\right)\quad &
\displaystyle \quad
\{x^a\,,\,p_b\}^{\ast\ast}=\delta^{a}{}_{b}\\[3mm] \displaystyle
\{x^a\,,\,\theta^\alpha\}^{\ast\ast}=-\frac{i}{2}
\frac{\epsilon^{abc}p_b(\sigma_c)^\alpha{}_{\beta}\theta^\beta}{p^2}&
\displaystyle\quad
\{\theta^\alpha\,,\,\theta^\beta\}^{\ast\ast}=-\frac{i}{2m}
\frac{p^{\alpha\beta}}{(-p^2)^{1/2}}\;.
\end{array} \label{36}
\end{equation}
Thus we result in $N=1$ superextension of the minimal anyon model with
monopole-like two-form (\ref{15}). The superparticle dynamics on the reduced
phase superspace is subject to mass-shell condition (\ref{11}) only
and the Hamiltonian reduces to
\begin{equation}
H=\frac{1}{2} e(\tau)(p^2+m^2)\;, \label{37}
\end{equation}
where $e(\tau)$ is a Lagrange multiplier. Because of the complicated
nonlinear structure of Dirac brackets (\ref{36}), it is a
nontrivial problem to obtain their Hilbert space operator realization.
That is why we choose another course to quantize this model.


\section{Quantization of the superanyon model}  \label{s4}

The quantization scheme of Sec.\ \ref{s2}, which was applied to the
anyon model with phase space $T^\ast({\Bbb R}^{1,2})\times {\cal L}$,
consists of combining the Dirac canonical quantization for the space-time
degrees of freedom with the geometric quantization for the curved inner
subspace. The efficiency of such an approach originated from the facts that
(i) the phase space is a product of two symplectic spaces;
(ii) the algebra of classical physical observables is spanned by functions
of the Poincar\'e generators;
(iii) the spin part of the Lorentz generators coincides with Berezin's
symbols for generators of the unitary representations $D^{|s|}_{\pm}$ of
$\overline{SU(1,1)}$. These features have natural generalizations in the
supersymmetric case, so the quantization scheme remains powerful too.

We have seen that the superanyon dynamics can be formulated, upon the
reduction with respect to the second class constraints (\thenumerone),
on the phase space $T^\ast({\Bbb R}^{1,2})\times{\cal L}^{1|1}$
which is a product of two symplectic (super) manifolds.
Similarly to the nonsupersymmetric case, all the classical observables are
functions of the $N=1$ super Poincar\'e generators (\ref{21}).
On $T^\ast({\Bbb R}^{1,2})\times{\cal L}^{1|1}$, the generators (\ref{21})
are constructed in terms of the space-time variables $x^a$, $p_a$ and
$osp(2|2)$-generators (\ref{33}).
The crucial point is that the $osp(2|2)$-generators prove to coincide with
Berezin's symbols of generators of an irreducible positive-weight
representation of
the superalgebra $osp(2|2)$\footnote{Strictly speaking,
we deal with so-called atypical representations of $osp(2|2)$
\cite{GradNiet96}.}
on superunit dick ${\cal L}^{1|1}$ [32\,--\,34].
That is why the quantization scheme described is well suited to the
superanyon model.
Let us start the quantization procedure with considering in more detail
the geometric quantization on the superunit disk.

Atypical unitary representations of the superalgebra $osp(2|2)$
can be realized in a ${\Bbb Z}_2$-graded space ${\cal O}_s$ of
antiholomorphic superfunctions over ${\cal L}^{1|1}$ of the form
\begin{equation}
f(\bar{z},\bar{\theta})=f_0(\bar{z})+\sqrt{s}\,\bar{\theta} f_1(\bar{z})
\qquad s>0\;, \label{39}
\end{equation}
where $f_{0,1}:{\cal L}\rightarrow {\Bbb C}$ are
ordinary antiholomorphic functions on the Lobachevsky plane.  A function
$f\in{\cal O}_s$ is said to be even if $f_1(\bar{z})= 0$ and odd if
$f_0(\bar{z})= 0$.  A representation ${\Bbb D}^s_+$ of $osp(2|2)$
of positive weight $s$
can be realized on ${\cal O}_s$ by choosing the  generators
in the form \cite{GradNiet96}
\begin{eqnarray}
\displaystyle
\hat{J}_a=-\bar{\xi}_a\bar\partial - \bar\partial\bar\xi_a
\left(s+\frac{1}{2}\bar\theta
\frac{\stackrel{\rightarrow}{\partial}}{\partial\bar\theta}\right)
\phantom{\pi_\alpha=\!}&\quad &
\displaystyle
\hat{B}=-s+\frac{1}{2}\bar\theta
\frac{\stackrel{\rightarrow}{\partial}}{\partial\bar\theta}\label{40}\\
\sqrt{2ms}\,\hat{\theta}^\alpha=\frac{1}{2}
\bar\theta(\bar{z}^\alpha\bar\partial+ 2s(\bar\partial \bar{z}^\alpha))
-\bar{z}^\alpha\frac{\stackrel{\rightarrow}{\partial}}{\partial\bar\theta}&
\quad &\displaystyle\hspace{-2.3pt}
\sqrt{\frac{2s}{m}}\,\hat{\pi}_\alpha=
\frac{i}{2}\bar\theta(\bar{z}_\alpha\bar\partial
+2s(\bar\partial \bar{z}_\alpha))+i\bar{z}_\alpha
\frac{\stackrel{\rightarrow}{\partial}}{\partial\bar\theta}\;.
\nonumber
\end{eqnarray}
The (anti) commutation relations for
$\hat{J}_a\,,\hat{B}\,,\hat{\theta}^\alpha$ and $\hat{\pi}_\alpha$ follow
from Eqs.\  (\ref{34}) by replacing $\{\;,\;\}^\ast\rightarrow
\frac{1}{i}[\;,\;]_{\mp}$ (anticommutator for two odd operators and
commutator in the rest cases).  With respect to the subalgebra $su(1,1)$ of
$osp(2|2)$, the representation splits into a sum of two irreducible unitary
representations of discrete series~${\Bbb D}^s_+=D^{s}_{+}
{\scriptstyle\bigoplus} D^{s+1/2}_{+}$. The even (odd) component of
$f\in{\cal O}_s$ transforms by
re\-pre\-sentation~${D^{s}_{+}\;(D^{s+1/2}_{+}}$).

The geometric quantization method
on ${\cal L}^{1|1}$ implies that the representation space is equipped with
the Hermitian two-form
\begin{equation}
\langle
f|g\rangle^s_{{\cal L}^{1|1}}=\int\nolimits_{{\cal L}^{1|1}}
{\rm d}\mu(z,\bar{z},\theta ,\bar{\theta})\,{\rm e}^{-s\Phi(z,\bar{z},\theta
,\bar{\theta})} \overline{f(\bar{z},\bar\theta)} g(\bar{z},\bar\theta)
\;,\label{41}
\end{equation}
where $f,g\in{\cal O}_s$, $\Phi(z,\bar{z},\theta
,\bar{\theta})$ is the K\"ahler superpotential ({\ref{32}) and
${\rm d}\mu(z,\bar{z},\theta ,\bar{\theta})$ is a Liouville supermeasure on
${\cal L}^{1|1}$.  Taking into account the definition of the closed
two-superform (\ref{31}), $\Omega\equiv {\rm d}r^A\Omega_{A\bar{B}} {\rm
d}r^{\bar{B}}\;,\; {\rm d}r^A\equiv ({\rm d}z,{\rm d}\theta)\;,\;{\rm
d}r^{\bar{A}}\equiv ({\rm d}\bar{z},{\rm d}\bar{\theta})$, one can calculate
the supermeasure explicitly \cite{BalantSchmittBarr88,Grad93}
\begin{equation}
{\rm d}\mu(z,\bar{z},\theta
,\bar{\theta})=-{\rm sdet}\left\|\Omega_{A\bar{B}}\right\|
\frac{{\rm d}z{\rm d}\bar{z}}{2\pi i}{\rm d}\theta {\rm d}\bar{\theta}=
-2\left(1-\frac{1}{2}\frac{\theta\bar{\theta}}{\zeta}
\right)\frac{{\rm d}z{\rm d}\bar{z}}{2\pi i}\frac{{\rm d}\theta
{\rm d}\bar{\theta}}{\zeta} \;.
\label{42}
\end{equation}
Accounting Eqs.\
(\ref{32}, \ref{42}), we integrate over the Grassmann variables in
(\ref{41}). Thus, the Hermitian form turns into
\begin{equation}
\langle f|g\rangle^s_{{\cal L}^{1|1}}= \langle
f_0|g_0\rangle_{{\cal L}}^s+ \langle f_1|g_1\rangle_{{\cal
L}}^{s+1/2}\;, \label{43}
\end{equation}
where $\langle\cdot|\cdot\rangle_{{\cal L}}^l$ is the
inner product for the representation space of $D^{l}_{+}$
\begin{displaymath}
\langle\varphi |\chi\rangle_{{\cal L}}^l=(2l-1)
\int\limits_{|z|<1}\frac{{\rm d}z{\rm d}\bar{z}}{2\pi i}\zeta^{2l-2}
\overline{\varphi (\bar{z})}\chi (\bar{z})\;.
\end{displaymath}
It is a matter of direct verification to prove that the generators
(\ref{40}) realize the irreducible unitary representation of
$osp(2|2)$.

Now we are in a position to construct the Hilbert space of the superanyon
states.
The space ${\cal H}$ of wave functions chosen in the form
\begin{equation}
F(p,\bar{z},\bar\theta)=F_0(p,\bar{z})+\sqrt{s}\,\bar\theta F_1(p,\bar{z})
\label{45}
\end{equation}
is naturally ${\Bbb Z}_2$-graded. The operator analogues for the classical
observables (\ref{21}) are defined by
\begin{equation}
\hat{{\cal J}}_a=-i\epsilon_{abc}p^b\frac{\partial}{\partial p_c}+\hat{J_a}
\qquad \hat{\cal P}_a=p_a \qquad
\hat{\cal Q}^{1}_\alpha=i(p\sigma)_{\alpha\beta}\hat{\theta}^\beta-
\hat{\pi}_\alpha\;.   \label{46}
\end{equation}
Owing to (\ref{ident}), the operator extensions for
(\ref{23}) can be chosen in the manner
\begin{equation}
\hat{\cal Q}^{2}_\alpha=-\frac{i}{m}(p\sigma)_{\alpha\beta}\hat{\pi}^\beta-
m\hat{\theta}_\alpha \qquad
\hat{{\cal K}}=1-2\bar\theta\frac{\stackrel{\rightarrow}{\partial}}{\partial
\bar\theta}\qquad \hat{{\cal Z}}=m \label{48}\;.
\end{equation}
Now, it is crucial to find the conditions, under which the operators
(\ref{46}) and (\ref{48})
realize a representation of the $N=2$ Poincar\'e superalgebra with central
charge. Straightforward calculations show that
the operators (\ref{46}) and (\ref{48}) satisfy almost all algebraic
relations (\ref{24}) but
\begin{eqnarray}&\displaystyle
[\hat{\cal Q}^I_\alpha\;,\;\hat{\cal Q}^J_\beta]_+=
2\delta^{IJ}p_{\alpha\beta}-2im\epsilon^{IJ}\epsilon_{\alpha\beta} &\nonumber
\\&\displaystyle
-\frac{1}{8ms}\bigl(p^2+m^2\bigr)\Bigl(4\delta^{IJ}\hat{J}_{\alpha\beta}
+i\epsilon^{IJ}\epsilon_{\alpha\beta}(4s-1+\hat{\cal K})\Bigr)&\\&
\displaystyle +\frac{1}{4ms}\bigl(4(p,\hat{J})+m(\hat{\cal K}-4s-1)\bigr)
\Bigl(\delta^{IJ}p_{\alpha\beta}-im\epsilon^{IJ}\epsilon_{\alpha\beta}\Bigr)
\;.&\nonumber\label{47}
\end{eqnarray}
Hence we conclude that the operators (\ref{46}, \ref{48})
form the superalgebra provided the wave
functions are subject to the equations
\begin{equation}
\begin{array}{c}\displaystyle
(p^2+m^2)F(p,\bar{z},\bar{\theta})=0 \\[2mm]  \displaystyle
(4(p,\hat{J})+m\hat{{\cal K}})F(p,\bar{z},\bar{\theta})=
m(4s+1)F(p,\bar{z},\bar{\theta})\,.
\end{array}
\label{49}
\end{equation}
These equations turn out to be {\it super Poincar\'e covariant}.
Moreover, the solutions of (\ref{49})
describe the superanyon doublet with the mass $m$ and the superspin
$s>0$. Accounting~(\ref{45}) the equations~(\ref{49}) are
reduced to
\begin{displaymath}
\begin{array}{ll}\displaystyle
\left\{
\begin{array}{l}
(p^2+m^2)F_0(p,\bar{z})=0 \\[2mm]
(p,\hat{J}^s)F_0(p,\bar{z})=msF_0(p,\bar{z})
\end{array}\right.&\displaystyle
\left\{
\begin{array}{l}
(p^2+m^2)F_1(p,z)=0 \\[2mm]
(p,\hat{J}^{s+1/2})F_1(p,\bar{z})=m(s+\frac{1}{2})F_1(p,\bar{z})\,.
\end{array}\right.
\end{array}
\label{50}
\end{displaymath}
where $\hat{J}^l=-\bar\xi_a\bar\partial-l\bar\partial\bar\xi_a\,,\,l=s,s+1/2$.
Comparing these equations with (\ref{17},~\ref{19}), one observes
that the even component of wave function $F(p,\bar{z},\bar\theta)$
describes the particle with spin $s$, whereas the odd one describes the
particle with spin $s+\frac{1}{2}$.

Finally, the space ${\cal H}^{m,s}$ of solutions to Eq. (\ref{49})
is naturally endowed with unique, modulo normalization,
super Poincar\'e and $osp(2|2)$ invariant inner product. It looks like
\begin{equation}
(F|G)=\int \frac{{\rm d}\vec{p}}{p^0}
\langle F|G \rangle^s_{{\cal L}^{1|1}}\qquad
p^0=\sqrt{\vec{p}^{\;2}+m^2}>0\;,  \label{51}
\end{equation}
where $\langle F|G \rangle^s_{{\cal L}^{1|1}}$ denotes the Hermitian
form (\ref{41}), (\ref{43}), $p^a\equiv(p^0,\vec{p}\;)$.
The generators (\ref{46}) realize the unitary irreducible representation of
the Poincar\'e superalgebra with spin $m$ and superspin $s>0$ in the space
${\cal H}^{m,s}$.  The case of $s<0$ can be treated in a similar way using
the doublet of representations $D_{-}^{-s}\bigoplus D_{-}^{-s-1/2}$.

It is remarkable that the construction proposed admits another interpretation
which is not related directly to geometric quantization.  It turns out that
the generators $\hat{\pi}_\alpha$ (or $\hat{\theta}^\alpha$) together
with the $U(1)$-charge $\hat{\cal K}$ realize a
representation of the deformed Heisenberg algebra (DHA) \cite{Vas89,Ply96}.
This follows from the identities
\begin{equation}
[\hat{\pi}_\alpha\;,\;\hat{\pi}_\beta]_{-}=\frac{m}{8s}\epsilon_{\alpha\beta}
(1+\nu \hat{\cal K}) \qquad
[\hat{\cal K}\;,\;\hat{\pi}_\alpha]_+=0\qquad \hat{\cal K}^2=1\,,\label{52}
\end{equation}
where
\begin{equation}\nu=4s-1\;.\end{equation}
The operators
$a^+=2\sqrt{2s/m}\,\hat{\pi}_1$ and $a=2\sqrt{2s/m}\,\hat{\pi}_0$ are
termed creation and annihilation operators, respectively,
$\nu$ is said to be deformation parameter. For $\nu=0$ (that corresponds
to supersemion $s=1/4$ \cite{SorTkVol89SorVol93})
the operators $\hat{\pi}_\alpha$ describe the usual
(undeformed) Heisenberg algebra. In the framework of the DHA approach
$\hat{\cal K}$ is known as Klein operator.

Now, one can reformulate the quantization in terms of the deformed oscillator
representation. The $osp(2|2)$-representation space ${\cal O}_s$ provides
us with a realization for the Fock space
of the deformed harmonic oscillator, the latter being
defined as
a linear space spanned by the vectors
$|0\rangle\,,\,|n\rangle=c_n(a^+)^n|0\rangle,\,n=1,2,\ldots$
($c_n$ is chosen
in such a way that $\langle n|n\rangle=1$). The Fock vacuum
$|0\rangle$ is defined by
\begin{equation}
a|0\rangle=0 \qquad\langle 0|0 \rangle=1 \qquad
\hat{\cal K}|0\rangle=|0\rangle\,.
\end{equation}
Since
\begin{displaymath}
a^+a|n\rangle=(n+\frac{\nu}{2}(1+(-1)^{n+1}))|n\rangle\;,
\end{displaymath}
the representation is unitary if $\nu>-1\;(s>0)$. The Klein operator
induces the ${\Bbb Z}_2$-graded structure in the Fock space
\begin{equation}
\hat{\cal K}|n\rangle=(-1)^n|n\rangle\,.
\end{equation}
The states $\{|2k\rangle,\,k=0,1,2\ldots\}$ form an orthonormal basis in the
even subspace, while the states $\{|2k+1\rangle,\,k=0,1,2\ldots\}$ in the
odd subspace.

The $osp(2|2)$ generators can be rewritten in terms of the DHA as follows:
\begin{equation}
\hat{J}_a=-\frac{2s}{m}(\sigma_a)_{\alpha\beta}\hat{\pi}^\alpha\hat{\pi}
^\beta \qquad
\hat{\theta}^\alpha=\frac{i}{m}\hat{K}\hat{\pi}^\alpha \qquad
\hat{B}=-\frac{1}{4}\hat{\cal K}(1-\nu\hat{\cal K})\;. \label{54}
\end{equation}
After that the quantization procedure can be performed in the same manner
we have already described.
Therefore, the superanyon doublet is naturally realized
in terms of the Fock space of the deformed
harmonic oscillator. For a fixed
momentum of the superparticle one can conceive the
spin-$s$ states live in the even subspace of the deformed Fock
space and the spin-$(s+\frac{1}{2})$ ones in the odd subspace.

It is worth pointing out that only
the supertranslations $\hat{\cal Q}^I_\alpha$
mix even and odd quantum states. The generators of the Poincar\'e algebra map
the even (odd) subspace of ${\cal H}$ on to itself
and this point was used in \cite{Ply96}
to realize the fractional spin one-particle states. The physical states
$F(p,\bar{z},\bar\theta)\in{\cal H}^{m,s}\subset{\cal H}$ were postulated
to be solutions of the following spinor equations
\begin{equation}
\left((p\sigma)_{\alpha\beta}\hat{\pi}^\beta+\epsilon m\hat{\pi}_\alpha
\right)F(p,\bar{z},\bar\theta)=0\qquad\epsilon=\pm\;\;.\label{55}
\end{equation}
One gets $F_1(p,\bar{z})=0$ for the solutions of (\ref{55}), while the even
component $F_0(p,\bar{z})$ describes the irreducible quantum dynamics
of the anyon with mass $m$ and spin $s=\epsilon(1+\nu)/4$.
It is the superanyon dynamics which makes use of all the
power of the DHA construction.

Sorokin, Tkach and Volkov \cite{SorTkVol89SorVol93}
showed that in three dimensions the dynamics of (super) particles with
(super) spin $1/4,3/4,5/4,\ldots$ can be naturally described by the use of the
usual undeformed oscillator representation ($\nu=0$). We have clarified
that the deformed Heisenberg algebra provides the description of dynamics
of arbitrary fractional (super)spin (super)particles.

\section{Conclusion  \label{s5}}
In the present paper we have constructed the classical and quantum dynamics
of superparticles with arbitrary fractional superspin in $D=1+2$ dimensions.
Our consideration was based on the use of $N=1$ supersymmetric
action functional (\ref{20}) which generalizes the anyon mechanical system
(\ref{1}) with the Lobachevsky plane in the role of spin space.
Thereby, Eq.\ (\ref{kok}) constitutes a supersymmetric generalization of the
Dirac monopole two-form, which is usually used for introducing
consistent couplings of $D=1+2$ particle to unconstrained backgroud fields
[17\,--\,20].  
It is believed that the superextension proposed offers a way to describe
$N=1$ superanyon dynamics in the presence of external superfields. Moreover,
the model (\ref{20}) possesses hidden invariance with respect to the $N=2$
Poincar\'e supergroup with the central charge whose on-shell value saturates
the BPS bound and, hence, corresponds to the shortening of $N=2$ massive
supermultiplets.

$N=2$ Poincar\'e supersymmetry is not the only hidden algebraic structure
originating in the model. In Hamilton approach, the system is
characterized by one first class and six second class constraints.
By restricting the dynamics to the surface of
second class constraints (25), one results in the formulation on reduced
phase space $T^\ast({\Bbb R}^{1,2})\times {\cal L}^{1|1}$, where
the K\"ahler supermanifold ${\cal L}^{1|1}=OSp(2|2)/U(1|1)$
is the minimal superextension of
the Lobachevsky space. The supergroup $OSp(2|2)$ is related
to the symplectic structure on ${\cal L}^{1|1}$
as the group of all superholomorphic
canonical transformations on ${\cal L}^{1|1}$.

Poincar\'e supersymmetry and $OSp(2|2)$ are closely related to each other,
both at the classical and quantum levels.
More precisely, the symplectic two-form (\ref{kok})
${\rm d}p_a\wedge {\rm d}x^a + s\Omega$ on the reduced phase space
is invariant under the $N=1$ supersymmetry transformations
on the mass-shell $p^2+m^2=0$, while, $\Omega$ remains unchanged
with respect to $OSp(2|2)$. That is why the super Poincar\'e
generators are built of the generators of $OSp(2|2)$ along with the
space-time co-ordinates and momenta.

The structure of the reduced phase space implies a natural technique
to quantize the model. It consists of combining the geometric
quantization on ${\cal L}^{1|1}$ and conventional quantization on
$T^\ast({\Bbb R}^{1,2})$. The $N=2$ Poincar\'e supersymmetry
turns out to be consistent provided imposing
the quantum equations of motion which single out the physical states of
superparticle.
Then the massive super Poincar\'e representation with the
superspin $s>0$ and the central charge equal to the mass $m$ is
realized on the superfields
transforming in the atypical representation of $osp(2|2)$
\cite{GradNiet96}, which splits, with respect to the subalgebra
$su(1,1)$ of $osp(2|2)$, in to
the doublet of discrete series representations
$D^s_+\bigoplus D^{s+1/2}_+$. Hence we obtain a direct superextension
of the well studied description of fractional spin states using the
representations $D^s_+$ [12\,--\,15, 21, 23].

The space of superparticle states with a fixed momentum
is shown to be embedded into
the Fock space of the deformed quantum oscillator. The deformation parameter
$\nu$ is related to the superspin by simple expression $\nu=4s-1\,(s>0)$.
This result generalizes some known constructions for anyons \cite{Ply96} and
(super) semions \cite{SorTkVol89SorVol93}.

We have studied the case of $N=1$ supersymmetric dynamics of anyons.
It would be of interest to extend the above consideration to the case
of $N$-extended Poincar\'e supersymmetry.
Here it is crucial to find an adequate analogue of the spin phase space
${\cal L}^{1|1}$. We hope to present respective constructions elsewhere.
\\[3mm]
{\bf Acknowledgements:} The authors are grateful to B.F. Samsonov for
interesting discussion and for drawing our attention to
Refs.~\cite{BalantSchmittBarr88,Grad93,GradNiet96}. The work of I.V.G.
is supported in part by the INTAS grant No. 93-2058-Ext.
\appendix
\renewcommand{\thesection}{}
\section{\hspace*{-0.65cm}Appendix A\hspace{0.8cm}Conventions \label{A}}
\setcounter{equation}{0}
\renewcommand{\theequation}{A.\arabic{equation}}
We define $D=1+2$ Minkowski metric $\eta_{ab}$ and Levi-Civita tensor
$\epsilon_{abc}$ as follows: $\eta_{ab}={\rm diag}(-,+,+)$ and
$\epsilon_{012}=-\epsilon^{012}=1$. Latin letters are used to denote
vector indices and Greek letters for spinor ones.
Due to the well-known isomorphism
$SO^{\uparrow}(1,2) \cong SU(1,1)/{\Bbb Z}_2$, the fundamental spinor
representation and its conjugate are defined by the transformation laws
$\psi_\alpha \rightarrow
N_\alpha{}^\beta\psi_\beta$, where $\alpha,\beta=0,1,$ and
$\bar{\psi}_{\dot \alpha} \equiv \overline{(\psi_\alpha)} \rightarrow
\overline{N}_{\dot \alpha}{}^{\dot \beta}{\bar \psi}_{\dot \beta}$
respectively.  Here $N \in SU(1,1)$ and $\overline{N}$ its complex conjugate
\begin{equation} \|N_\alpha{}^\beta\|= \left( \begin{array}{cc} a & b \\ \bar
	       b & \bar a \end{array}\right) \qquad |a|^2-|b|^2=1\;.
\end{equation}
The spinor representations are equivalent, since $SU(1,1)$ possesses not
only invariant spinor antisymmetric metric
$\epsilon_{\alpha\beta}=-\epsilon_{\beta\alpha}=-\epsilon^{\alpha\beta}$
$(\epsilon_{01}=1)$ and its conjugate, which are used for raising and
lowering spinor indices by the rule
$\psi_\alpha=\epsilon_{\alpha\beta}\psi^\beta,\,
\psi^\alpha=\epsilon^{\alpha\beta}\psi_\beta$,
but also the invariant tensor with mixed indices
\begin{equation}
g_{\alpha{\dot \alpha}}=
\left( \begin{array}{cc}
		    1 & 0 \\
		    0 & -1
       \end{array}\right)
\end{equation}
that allows to convert dotted spinor indices into undotted ones in the manner
$\bar{\psi}_\alpha = g_\alpha{}^{\dot \alpha} \bar{\psi}_{\dot \alpha}$,
$\bar{\psi}_{\dot \alpha} = g_{\dot \alpha}{}^\alpha \bar{\psi}_\alpha$,
where $g_\alpha{}^{\dot \alpha} =\epsilon^{\dot \alpha \dot \beta}
g_{\alpha \dot \beta}$ and
$g_{\dot \alpha}{}^\alpha =\epsilon^{\alpha \beta} g_{\beta \dot \alpha}$.
This makes it possible to use undotted spinors only.

Spinors may be subject to a covariant reality condition of the form
\begin{equation}
\bar{\psi}_\alpha =\Delta \psi_\alpha  \quad \Longleftrightarrow \quad
\bar{\psi}^\alpha = - \Delta \psi^\alpha \qquad |\Delta| =1,
\end{equation}
for some parameter $\Delta$.
We choose $\Delta =1$ for the odd co-ordinates $\theta^\alpha$ of
$N=1,D=3$ superspace.

The Dirac matrices are chosen in the form
\begin{equation}
(\sigma_0)_{\alpha\beta}=
\left( \begin{array}{cc}
		    0 & 1 \\
		    1 & 0
       \end{array}\right)\quad
(\sigma_1)_{\alpha\beta}=
\left( \begin{array}{cc}
		    1 & 0 \\
		    0 & 1
       \end{array}\right)\quad
(\sigma_2)_{\alpha\beta}=
\left( \begin{array}{cc}
		   -i  & 0 \\
		    0 & i
       \end{array}\right)
\label{a1}
\end{equation}
\begin{displaymath}
(\sigma_a)_{\alpha\gamma}(\sigma_b)^{\gamma}{}_\beta=
i\epsilon_{abc}(\sigma^c)_{\alpha\beta}-\eta_{ab}\epsilon_{\alpha\beta}
\end{displaymath}
such that the matrices
$(\sigma_a)_{\alpha \dot \alpha} = g_{\dot \alpha}{}^\beta
(\sigma_a)_{\alpha \beta}$ are Hermitian.
The double-sheeted covering map $\pi :SU(1,1)\rightarrow SO^\uparrow (1,2)$
mentioned is constructed with the help of the $\sigma$-matrices by
associating with an element $N = \|N_\alpha{}^\beta\| \in SU(1,1)$ its
image
$\Lambda(N)=\|\Lambda^a{}_b\|\in SO^\uparrow (1,2)$,
in the connected component of the identity
of the Lorentz group, defined by
\begin{equation}
\Lambda^b{}_a(\sigma_b)_{\alpha\dot \alpha}=N_\alpha{}^\beta
\overline{N}_{\dot \alpha}{}^{\dot \beta}
(\sigma_a)_{\beta \dot \beta}\;.
\label{a2}
\end{equation}

We follow Berezin's conventions for superforms \cite{Berezin}.
The Grassmann parity $\epsilon (\Omega)$ in
a superalgebra of exterior superforms
is defined by requiring
that (i) the Grassmann parity of an even (odd) $0$-form is equal to
0 (1); (ii) the Grassmann parity of exterior differential is equal to 1,
$\epsilon ({\rm d}\Omega )=\epsilon (\Omega )+1$. Thus, if $r^A$ are
coordinates on a supermanifold of parity $\epsilon_A$, then
$r^Ar^B=(-1)^{\epsilon_A\epsilon_B}r^Br^A\;,\;
{\rm d}r^Ar^B=(-1)^{\epsilon_B(\epsilon_A+1)}r^B{\rm d}r^A\;,\;
{\rm d}r^A{\rm d}r^B=(-1)^{(\epsilon_A+1)(\epsilon_B+1)}{\rm d}r^B{\rm d}r^A$.
Finally, the Leibniz rule looks like
${\rm d}(\Omega_1\Omega_2)={\rm d}(\Omega_1)\Omega_2+(-1)^{\epsilon(\Omega_1)}
\Omega_1{\rm d}\Omega_2$.

\section{\hspace*{-6.5mm}Appendix B\protect\\
\hspace*{-6.5mm}Lobachevsky plane as a homogeneous space \label{B}}
\setcounter{equation}{0}\renewcommand{\theequation}{B.\arabic{equation}}
Here we describe a ``manifestly Lorentz-covariant"  realization of
Lobachevsky plane ${\cal L}=SU(1,1)/U(1)$
as a homogeneous space of $SO^\uparrow(1,2)$. This realization is used
throughout the paper.
${\cal L}$ is identified with a unit open disc in a complex plane,
${\cal L} \cong\{z\in{\Bbb C},|z|<1\}$. The proper orthochronous Lorentz
group $SO^\uparrow(1,2)\cong SU(1,1)/{\Bbb Z}_2$ acts on $\cal L$ by
fractional linear transformations
\begin{equation} N:\;\;z\rightarrow
z^{\prime}=\frac{az-b}{\bar a-\bar{b} z} \qquad \qquad N \in
SU(1,1)\;.\label{b1} \end{equation} One can rewrite Eq.\ (\ref{b1}) in a
manifestly covariant form by introducing the two-component twistor-like
objects
\begin{equation}
z^\alpha\equiv(1,z) \qquad \quad \bar{z}^\alpha\equiv(\bar{z},1)
\label{mist}
\end{equation}
transforming by the law
\begin{equation}
N:\ z^{\alpha}\to z^{\alpha\,\prime} = \left(
\frac{\partial z^\prime}{\partial z}\right )^{1/2} N^{-1}{}_\beta {}^\alpha
z^\beta\qquad \bar{z}^\alpha\rightarrow \bar{z}^{\alpha\,\prime}= \left(
\frac{\partial \bar{z}^\prime}{\partial \bar{z}}\right )^{1/2} N^{-1}{}_\beta
{}^\alpha \bar{z}^\beta\;,
\label{b2} \end{equation}
or, in infinitesimal form,
\begin{equation}
\delta z=\frac{i}{2}\omega_{\alpha\beta} z^{\alpha} z^{\beta}\qquad
\delta \bar{z}=-\frac{i}{2}\omega_{\alpha\beta} \bar{z}^{\alpha}
\bar{z}^{\beta}\,,\label{b3}
\end{equation}
where $\omega_{\alpha\beta}\equiv(\omega^a\sigma_a )_{\alpha\beta}$
are the parameters of infinitesimal Lorentz transformations.
As it is seen, each of
$z^\alpha$ and $\bar{z}^\alpha$ transforms simultaneously
as a $D=3$ Lorentz spinor and a tensor field on ${\cal L}$.
Using $z^\alpha$ and $\bar{z}^\alpha$ we may construct the following
vector densities
\begin{eqnarray}
&&\zeta_a\equiv -(\sigma_a )_{\alpha\beta} z^\alpha
\bar{z}^{\beta}=-(1+z\bar{z},z+\bar{z},i(z-\bar{z}))\label{b4}\\
&&\xi_a\equiv -\frac{1}{2}(\sigma_a )_{\alpha\beta} z^\alpha
              z^{\beta}=-\frac{1}{2}(2z,1+z^2,i(z^2-1))\qquad
\bar{\xi}_a\equiv (\xi_a)^*
\label{b5}
\end{eqnarray}
and the scalar density
\begin{equation}
\zeta\equiv\epsilon_{\alpha\beta}z^\alpha\bar{z}^\beta=1-z\bar{z}\qquad
\zeta^a\zeta_a=-2\xi^a\bar{\xi}_a=-\zeta^2\label{b7}
\end{equation}
as well.
The following identity
\begin{equation}
4\frac{\xi_a\bar\xi_b}{\zeta^2}\equiv i\epsilon_{abc}n^c+n_an_b+\eta_{ab}
\qquad n_a\equiv\frac{\zeta_a}{\zeta} \label{B8}
\end{equation}
is useful in practice.
The chief advantage of the technique described consists in the
fact that $z^\alpha$ and $\bar{z}^\alpha$ are the only independent
tensor-like fields associated with the homogeneous space structure
on $\cal L$. Our treatment here follows Ref.\ \cite{KuzLyakSeg95}
where objects like $z^\alpha$ were introduced on two-sphere $S^2$.


\end{document}